\documentclass[10pt, a4paper]{article}

\usepackage[utf8]{inputenc}
\usepackage[T1]{fontenc}
\usepackage{lmodern,textcomp}
\usepackage[english]{babel}
\usepackage{csquotes}

\usepackage[top=4cm,bottom=4cm,left=3.5cm,right=3.5cm]{geometry}

\usepackage{eurosym,xspace}
\usepackage[nottoc]{tocbibind}
\usepackage{graphicx,subfig,float}
\usepackage[multiple]{footmisc}
\usepackage{authblk}
\usepackage[backend=bibtex8, citestyle=numeric-comp, maxbibnames=99, sorting=nyt,
			sortcites=true, firstinits=true, sorting=none]{biblatex}
\input{arxiv.def}

\usepackage{amsmath,amsfonts,amssymb}

\usepackage{maths_min}

\usepackage[pdftex, colorlinks=true, breaklinks=true,
	urlcolor=blue, linkcolor=blue, citecolor=red,
	pdfcreator={PDFLaTeX}, pdfauthor={Harold Erbin}
]{hyperref}
\usepackage[all]{hypcap}

\hypersetup{
	pdftitle={Deciphering and generalizing Newman–Janis–Demiański algorithm},
}

\numberwithin{equation}{section}

\newcommand{\email}[1]{\thanks{\href{mailto:#1}{\texttt{#1}}}}

\title{Deciphering and generalizing Demiański–Janis–Newman algorithm}

\author[1]{Harold Erbin\email{erbin@lpthe.jussieu.fr}}
\affil[1]{Sorbonne Universités, UPMC Univ Paris 06, UMR 7589, LPTHE, F-75005, Paris, France}
\affil[1]{CNRS, UMR 7589, LPTHE, F-75005, Paris, France}

\begin{document}

\maketitle

\begin{abstract}
In the case of vanishing cosmological constant, Demiański has shown that the Janis–Newman algorithm can be generalized in order to include a NUT charge and another parameter $c$, in addition to the angular momentum.
Moreover it was proved that only a NUT charge can be added for non-vanishing cosmological constant.
However despite the fact that the form of the coordinate transformations was obtained, it was not explained how to perform the complexification on the metric function, and the procedure does not follow directly from the usual Janis–Newman rules.
The goal of our paper is threefold: explain the hidden assumptions of Demiański's analysis, generalize the computations to topological horizons (spherical and hyperbolic) and to charged solutions, and explain how to perform the complexification of the function.
In particular we present a new solution which is an extension of the Demiański metric to hyperbolic horizons.
These different results open the door to applications in (gauged) supergravity since they allow for a systematic application of the Demiański–Janis–Newman algorithm.
\end{abstract}

\newpage

\hrule
\pdfbookmark[1]{\contentsname}{toc}
\tableofcontents
\bigskip
\hrule

\section{Introduction}

The Janis–Newman (JN) algorithm is a procedure which transforms a seed configuration to another configuration with more conserved charges~\cite{Newman:1965:NoteKerrSpinningParticle, Newman:1965:MetricRotatingCharged, Adamo:2014:KerrNewmanMetricReview}.
Most of the applications of this algorithm have been dedicated to obtaining rotating solutions without cosmological constant from static ones~\cite{Herrera:1982:ComplexificationNonrotatingSphere, Drake:1997:ApplicationNewmanJanisAlgorithm, Drake:2000:UniquenessNewmanJanisAlgorithm, Yazadjiev:2000:NewmanJanisMethodRotating, Ibohal:2005:RotatingMetricsAdmitting}, but it has been shown by Demiański (and partially by Newman) that other parameters can be added~\cite{Demianski:1966:CombinedKerrNUTSolution, Demianski:1972:NewKerrlikeSpacetime}, even in the presence of a cosmological constant.
In what follows we call this generalized procedure the Demiański–Janis–Newman (DJN) algorithm.\footnotemark{}
\footnotetext{Demiański's metric has been generalized in~\cite{Patel:1978:RadiatingDemianskitypeSpacetimes, Krori:1981:ChargedDemianskiMetric, Patel:1988:RadiatingDemianskitypeMetrics}.}

The (D)JN algorithm relies on a complex coordinate transformation.
The original formulation uses the Newman–Penrose tetrad formalism, but a simpler prescription that can be used to transform directly the metric has been proposed by Giampieri~\cite{Giampieri:1990:IntroducingAngularMomentum, Erbin:2015:JanisNewmanAlgorithmSimplifications}.
All our computations are done using this last approach since both formalisms are totally equivalent.

In order to find the most general complex transformation, Demiański starts from a metric ansatz (with one unknown function) and introduces two arbitrary $\theta$-dependent functions in the complex transformation~\cite{Demianski:1972:NewKerrlikeSpacetime}.
After application of the JN algorithm,\footnotemark{} solving Einstein equations provides a generic form for the allowed transformations.
\footnotetext{Note that JN algorithm is off-shell: in general it does not preserve the equations of motion and it can be performed on any field configuration, not only on solutions.}
In particular the latter depend on three parameters (including angular momentum and NUT charge) for vanishing cosmological constant, and only on the NUT charge for non-vanishing cosmological constant.
Moreover determining the stationary metric functions\footnotemark{} from Einstein equations ensures that there is no ambiguity in the transformation, as opposed to the usual algorithm, and this allows to focus on the determination of the complex transformation.
\footnotetext{We call a "seed/stationary metric function" a function that appears in the seed/stationary metric. The term "stationary" is used to describe the metric resulting from the DJN algorithm, which generically is non-static.}
This approach can be particularly fruitful in cases where one does not know precisely the rules of the algorithm (such as in higher dimensions~\cite{Erbin:2015:FivedimensionalJanisNewmanAlgorithm}).

Demiański's paper~\cite{Demianski:1972:NewKerrlikeSpacetime} is short and results are extremely condensed.
A first goal of our paper is to expose the full technical details of the computations.
In particular we uncover underlying assumptions on the form of the metric function and correct an error in his formula (14) (already found in~\cite{Quevedo:1992:ComplexTransformationsCurvature}).

One of the obvious generalizations is the inclusion of a gauge field which is needed to obtain (electrically) charged solutions.
It appears that the analysis remains unchanged, the Maxwell equations being also integrable within Demiański's ansatz.
This solution was already found in~\cite{Krori:1981:ChargedDemianskiMetric} for spherical horizon topology, but we extend it to hyperbolic topology and we demonstrate how to perform the full computation using the DJN algorithm, as a first step towards possible generalizations to other cases.
This computation has been made possible by the recent discovery of the way to apply the JN algorithm to gauge fields~\cite{Erbin:2015:JanisNewmanAlgorithmSimplifications} (see also~\cite{Keane:2014:ExtensionNewmanJanisAlgorithm, Adamo:2014:KerrNewmanMetricReview} for other approaches).

Another improvement of the DJN algorithm that results of our analysis is the generalization of all formulas to topological horizons.
In particular all existing formulas can be straightforwardly generalized to the case of hyperbolic horizons,\footnotemark{} and we prove all formulas by solving explicitly Einstein equations.
\footnotetext{We do not treat the case of flat horizon but this could be obtained from some easy reparametrization.}

We also comment on the group properties that some of the DJN transformations possess.
This observation can be useful for chaining several transformations, therefore adding charges to a solution that is already non-static (for example adding rotation to a solution that already contains a NUT charge).
More importantly this provides a setting where the algorithm \emph{does} preserve the Einstein equations.

The main drawback in determining the stationary functions from Einstein equations is that one does not get the rules for deriving them from the seed functions.
For example it is not obvious how to obtain the stationary function from the static one in Demiański's paper.
If there were no way to obtain the functions by complexification then the DJN algorithm would be of limited interest as it could not be exported to other cases (except if one is willing to solve Einstein–Maxwell equations, which is not the goal of a solution generating technique).
We demonstrate that the transformation can be achieved by a complexification of the mass together with a shift of the horizon curvature
\begin{equation}
	m = m' + i \kappa\, n, \qquad
	 \kappa \longrightarrow \kappa - \frac{4\Lambda}{3}\, n^2,
\end{equation} 
$n$ being the NUT charge, establishing that Demiański's transformations can be interpreted as an extension of the usual JN algorithm.
Note that it is necessary to introduce $\kappa$ in order to perform the transformation, even if one is interested only in spherical topology with $\kappa = 1$.
Such a complex combination is quite natural from the point of view of the Plebański–Demiański solution~\cite{Plebanski:1975:ClassSolutionsEinsteinMaxwell, Plebanski:1976:RotatingChargedUniformly}.
Complexification of parameters in the context of a solution generating technique was also done by Quevedo~\cite{Quevedo:1992:ComplexTransformationsCurvature, Quevedo:1992:DeterminationMetricCurvature}.

We end the introduction by describing our ansatz.
We consider the most general seed metric for which $(\theta, \phi)$-section are $2$-dimensional maximally symmetric spaces (it can be the sphere $S^2$ or the hyperboloid $H^2$) and with only radial functions.
Similarly the gauge field contains only one unknown radial function and is purely electric.
The DJN algorithm generates a stationary metric coupled to a gauge field for a total of five unknown (independent) functions.
We provide several formulas in $(u, r)$ and $(t, r)$ coordinates that are suitable for any application of the DJN algorithm.\footnotemark
\footnotetext{We stress that at this stage these formulas do not satisfy Einstein equations, they are just a proxy to simplify later computations.}
Similar formulas for subcases have been obtained in~\cite{Whisker:2008:BraneworldBlackHoles, Drake:2000:UniquenessNewmanJanisAlgorithm, AzregAinou:2014:StaticRotatingConformal, AzregAinou:2014:GeneratingRotatingRegular}.
All these computations are gathered in a Mathematica file (available on demand) which includes an implementation of Einstein–Maxwell equations.
We insist on the fact that all these results can also be derived from the tetrad formalism.

The paper is organized as follows.
In section~\ref{sec:ansatz} we set up our ansatz for the seed metric and gauge field.
Then in section~\ref{sec:generalized-jna} we apply the the DJN algorithm on the previous ansatz.
In particular we explain how to generalize it to topological horizons.
In section~\ref{sec:charged-demianski} we solve Einstein–Maxwell equations when there is one unknown function in the metric.
In section~\ref{sec:group-properties} we make a brief comment on the group formed by a subclass of the JN transformations.
Finally in section~\ref{sec:complexification} we explain how to perform the complexification of the metric function in order to recover the results obtained from Einstein–Maxwell equations.
We provide two appendices.
In appendix~\ref{app:original-demianski} we recall original Demiański's solution, while in appendix~\ref{app:relaxing-assumptions} we relax simplifying assumptions made in the main sections.

\section{Setting up the ansatz}
\label{sec:ansatz}

In this section we recall Einstein–Maxwell equations before describing the ansatz we will use for the Janis–Newman algorithm in section~\ref{sec:generalized-jna}.
The equations will be solved for a simple case in section~\ref{sec:charged-demianski}.

Equations of motion for Einstein–Maxwell gravity with cosmological constant $\Lambda$ read
\begin{equation}
	\label{top-down:eq:einstein-maxwell-eom}
	G_{\mu\nu} + \Lambda g_{\mu\nu} = 2\, T_{\mu\nu}, \qquad
	\grad_\mu F^{\mu\nu} = 0,
\end{equation} 
where $F = \dd A$ and the stress–energy tensor for the electromagnetic gauge field $A_\mu$ is
\begin{equation}
	T_{\mu\nu} = F_{\mu\rho} \tens{F}{_\nu^\rho} - \frac{1}{4}\, g_{\mu\nu} F^2.
\end{equation} 
In our conventions the spacetime signature is mostly plus and we have set Einstein's constant $\kappa$ to $1$.

The static electromagnetic one-form is taken to be
\begin{equation}
	\label{top-down:eq:static-gauge-field}
	A(r) = f_A(r)\, \dd t.
\end{equation} 
This ansatz is purely electric since only the time component is non-zero.

The static metric ansatz in coordinates $(t, r, \theta, \phi)$ reads
\begin{equation}
	\label{top-down:eq:static-metric}
	\dd s^2 = - f_t(r)\, \dd t^2 + f_r(r)\, \dd r^2 + f_\Omega(r)\, \dd\Omega^2.
\end{equation} 
One of the functions is redundant since we are free to redefine the radial coordinate.

The $(\theta, \phi)$ sections correspond to $2$-dimensional maximally symmetric spaces, which are the sphere $S^2$, the euclidean plane $\R^2$ and the hyperboloid $H^2$ respectively for positive, vanishing and negative curvature~\cite{AlonsoAlberca:2000:SupersymmetryTopologicalKerrNewmannTaubNUTaDS}.
Defining $\kappa$ as the sign of the surface curvature, the uniform metric $\dd\Omega^2$ is given by
\begin{equation}
	\dd \Omega^2 = \dd\theta^2 + H(\theta)^2\, \dd \phi^2
\end{equation} 
with
\begin{equation}
	H(\theta) =
	\begin{cases}
		\sin \theta & \kappa = 1, \\
		1 & \kappa = 0, \\
		\sinh \theta & \kappa = -1.
	\end{cases}
\end{equation}
In the rest of the paper we focus on $\kappa = \pm 1$.

Introducing the null coordinates $u$ through the change of coordinates
\begin{equation}
	\label{top-down:change:null-diff}
	\dd t = \dd u + \sqrt{\frac{f_r}{f_t}}\, \dd r,
\end{equation} 
the static metric \eqref{top-down:eq:static-metric} becomes
\begin{equation}
	\label{top-down:eq:static-metric-ur}
	\dd s^2 = - f_t\, \dd u^2 - 2 \sqrt{f_t f_r}\; \dd u \dd r + f_\Omega\, \big( \dd\theta^2 + H^2\, \dd \phi^2\big),
\end{equation} 
while the gauge field \eqref{top-down:eq:static-gauge-field} is found to be
\begin{equation}
	A = f_A \left(\dd u + \sqrt{\frac{f_r}{f_t}}\, \dd r \right).
\end{equation} 
Since the component $A_r$ depends only on $r$ it can be removed by a gauge transformation~\cite{Erbin:2015:JanisNewmanAlgorithmSimplifications} such that
\begin{equation}
	\label{top-down:eq:static-gauge-field-ur}
	A = f_A\, \dd u.
\end{equation}

\section{Generalized Janis–Newman algorithm}
\label{sec:generalized-jna}

In this section we apply the Janis–Newman algorithm to the ansatz of the previous section.
Using arbitrary functions for the complex transformation and for the functions inside the metric, we obtain a very general ansatz; then we will solve Einstein–Maxwell equations in the next section in order to find their forms.
We will directly use Giampieri's prescription~\cite{Giampieri:1990:IntroducingAngularMomentum, Erbin:2015:JanisNewmanAlgorithmSimplifications} in order to avoid the introduction of tetrads and the computation of the contravariant components of the metric and of the gauge field, but we stress that it is fully equivalent to the tetrad formalism~\cite{Giampieri:1990:IntroducingAngularMomentum, Erbin:2015:JanisNewmanAlgorithmSimplifications}.
Reviews and simpler applications can be found in~\cites{Adamo:2014:KerrNewmanMetricReview, Drake:2000:UniquenessNewmanJanisAlgorithm, Erbin:2015:JanisNewmanAlgorithmSimplifications}[sec.~5.4]{Whisker:2008:BraneworldBlackHoles}.

\subsection{Janis–Newman transformation}
\label{sec:generalized-jna:jna-transf}

The Janis–Newman algorithm can be summarized as the following sequence of steps:
\begin{enumerate}
	\item Start with a seed metric in $(u, r)$ coordinates.
	
	\item Let the coordinates $u$ and $r$ become complex.
	
	\item Replace the functions initially inside the metric by other (real) functions depending on $r$ and its conjugate, such that the metric remains real.
	
	\item Make a change of coordinates $(r, u) \to (r', u')$, the new coordinates being real.
	
	\item Apply Giampieri's ansatz to recover a real metric.
\end{enumerate}

The complex change of coordinates is given by\footnotemark{}~\cite{Demianski:1972:NewKerrlikeSpacetime}
\footnotetext{Similar transformations have been studied by Talbot~\cite{Talbot:1969:NewmanPenroseApproachTwisting}.}
\begin{equation}
	\label{top-down:change:jna-coord}
	r = r' + i\, F(\theta), \qquad
	u = u' + i\, G(\theta),
\end{equation} 
\footnotetext{In his paper~\cite{Demianski:1972:NewKerrlikeSpacetime} Demiański considers functions that depend on $\theta$ and $\phi$, but he drops the $\phi$-dependence at an intermediate step.
Because we want to keep a $\group{U}(1)_\phi$ isometry we will follow him and ignore any $\phi$ dependence.}
Usually these functions are taken to be
\begin{equation}
	F(\theta) = - a \cos \theta, \qquad
	G(\theta) = a \cos \theta,
\end{equation} 
but here they are kept general and the most general transformation will be determined by Einstein equations.

As given by\footnotemark{}
\begin{equation}
	\dd r = \dd r' + i\, F'(\theta)\, \dd\theta, \qquad
	\dd u = \dd u' + i\, G'(\theta)\, \dd\theta,
\end{equation} 
the differentials of the coordinates are complex which is not coherent with having a real metric.
\footnotetext{The prime on $F$ and $G$ denoting the differentiation with respect to $\theta$. In general primes on functions denote derivative with respect to its argument – which is $\theta$ here –, while primes on $u$ and $r$ indicates new variables.}
The (generalized) Giampieri's ansatz consists in the replacement
\begin{equation}
	i\, \dd \theta = \sqrt{g^\Omega_{\phi\phi}}\, \dd\phi
		= H(\theta)\, \dd \phi,
\end{equation} 
where the RHS is given by comparison of the final result with the tetrad formalism~\cite{Giampieri:1990:IntroducingAngularMomentum, Erbin:2015:JanisNewmanAlgorithmSimplifications} (see also~\cite{Ferraro:2014:UntanglingNewmanJanisAlgorithm, Nawarajan:2016:CartesianKerrSchildVariation} for related motivations).
As a consequence the transformation of the differentials is
\begin{equation}
	\label{top-down:change:jna-diff}
	\dd r = \dd r' + F'(\theta) H(\theta)\, \dd \phi, \qquad
	\dd u = \dd u' + G'(\theta) H(\theta)\, \dd \phi.
\end{equation} 

Finally the four functions
\begin{equation}
	f_i = f_i(r), \qquad
	f_i = \{ f_t, f_r, f_\Omega, f_A \}
\end{equation} 
are replaced by
\begin{equation}
	\tilde f_i = \tilde f_i(r, \bar r), \qquad
	\tilde f_i = \{ \tilde f_t, \tilde f_r, \tilde f_\Omega, \tilde f_A \}.
\end{equation} 
There are only three conditions that these functions satisfy
\begin{equation}
	\label{top-down:eq:conditions-tilde-f}
	\tilde f_i(r, \bar r) = \tilde f_i \big(r', F(\theta) \big), \qquad
	\tilde f_i(r', 0) = f_i(r'), \qquad
	\tilde f_i(r, \bar r) \in \R.
\end{equation} 
The first relation means that the dependence in $\theta$ is solely contained in the functional dependence of $F(\theta)$, as it is evident from the explicit form of $r$ and $\bar r$.\footnotemark{}
Note that here we do not try to get the functions $\tilde f_i$ from the complexification of the static functions~\cite{Demianski:1972:NewKerrlikeSpacetime}; this is the topic of section~\ref{sec:complexification}.
\footnotetext{This condition is not explicit in Demiański's paper~\cite{Demianski:1972:NewKerrlikeSpacetime} but it is useful to take it into account in the computations.}

As a consequence of \eqref{top-down:eq:conditions-tilde-f} the $\theta$-derivative of $\tilde f_i$ reads
\begin{equation}
	\pd_\theta \tilde f_i = F'\, \pd_F \tilde f_i
\end{equation} 
such that it is sufficient to obtain the dependence of $\tilde f_i$ in terms of $F$.

\subsection{Stationary metric}

Applying the transformations \eqref{top-down:change:jna-coord} and \eqref{top-down:change:jna-diff} and replacing the functions, the resulting stationary metric in Eddington–Finkelstein coordinates is (removing the primes on $u$ and $r$ to lighten the notations)
\begin{equation}
	\label{top-down:eq:rotating-metric:ur}
	\dd s^2 = - \tilde f_t (\dd u + \alpha\, \dd r + \omega H\, \dd\phi )^2
		+ 2 \beta\, \dd r \dd \phi
		+ \tilde f_\Omega \big(\dd\theta^2 + \sigma^2 H^2 \dd\phi^2 \big)
\end{equation} 
where we defined the quantities
\begin{equation}
	\omega = G' + \sqrt{\frac{\tilde f_r}{\tilde f_t}}\, F', \qquad
	\sigma^2 = 1 + \frac{\tilde f_r}{\tilde f_\Omega}\, F'^2, \qquad
	\alpha = \sqrt{\frac{\tilde f_r}{\tilde f_t}}, \qquad
	\beta = \tilde f_r\, F' H.
\end{equation}

The transformation
\begin{equation}
	\label{top-down:change:bl-diff}
	\dd u = \dd t - g(r) \dd r, \qquad
	\dd \phi = \dd \phi' - h(r) \dd r
\end{equation} 
can be used to set the coefficient $g_{ur}$ and $g_{r\phi}$ to zero and to cast the metric in Boyer–Lindquist (BL) coordinates.
The solution to these two conditions is
\begin{equation}
	\label{top-down:change:bl-gh}
	g(r) = \frac{\sqrt{\big(\tilde f_t \tilde f_r \big)^{-1}}\, \tilde f_\Omega - F' G'}{\Delta}, \qquad
	h(r) = \frac{F'}{H(\theta) \Delta}
\end{equation} 
with
\begin{equation}
	\Delta = \frac{\tilde f_\Omega}{\tilde f_r} + F'^2
		= \frac{\tilde f_\Omega}{\tilde f_r}\, \sigma^2.
\end{equation} 
We stress that the functions $g$ and $h$ cannot depend on $\theta$, otherwise the change of variables \eqref{top-down:change:bl-diff} is not integrable.
It is thus necessary to check for given functions $\tilde f_i, F$ and $G$ that all the $\theta$-dependence cancel.

Finally the metric in $(t, r)$ coordinates can be written (removing the prime on $\phi$)
\begin{equation}
	\label{top-down:eq:rotating-metric:tr}
	\dd s^2 = - \tilde f_t (\dd t + \omega H\, \dd\phi )^2
		+ \frac{\tilde f_\Omega}{\Delta}\, \dd r^2
		+ \tilde f_\Omega \big(\dd\theta^2 + \sigma^2 H^2 \dd\phi^2 \big).
\end{equation} 

All expressions are invariant under the transformation
\begin{equation}
	(F, G, \phi) \longrightarrow - (F, G, \phi).
\end{equation} 

\subsection{Gauge field}

Applying the DJN transformations \eqref{top-down:change:jna-diff} to the gauge field \eqref{top-down:eq:static-gauge-field-ur}
\begin{equation}
	A = f_A\, \dd u
\end{equation}
gives\footnotemark{} (removing the prime on $u$)
\footnotetext{This may also be derived from the tetrad formalism~\cite{Keane:2014:ExtensionNewmanJanisAlgorithm, Adamo:2014:KerrNewmanMetricReview, Erbin:2015:JanisNewmanAlgorithmSimplifications}.}
\begin{equation}
	\label{top-down:eq:rotating-gauge-field:ur}
	A = \tilde f_A\, (\dd u + G' H\, \dd \phi)
\end{equation} 
 
Using the explicit formula \eqref{top-down:change:bl-gh}, the previous expression becomes in Boyer–Lindquist coordinates
\begin{equation}
	\label{top-down:eq:rotating-gauge-field:tr}
	A = \tilde f_A\, \left( \dd t - \frac{\tilde f_\Omega}{\sqrt{\tilde f_t \tilde f_r}\, \Delta}\, \dd r + G' H\, \dd \phi \right).
\end{equation} 
Here the function
\begin{equation}
	A_r = - \frac{\tilde f_A \tilde f_\Omega}{\sqrt{\tilde f_t \tilde f_r}\, \Delta}
\end{equation} 
may depend on $\theta$ in which case it would not be possible to remove it by a gauge transformation.\footnotemark{}
\footnotetext{Note that in several examples where BL coordinates exist, $A_r$ depends only on $r$ and hence is removable by a gauge transformation, such that this situation seems to be generic.}

\section[Einstein–Maxwell equations: Charged topological Demiański's solution]{Einstein–Maxwell equations: Charged topological\\ Demiański's solution}
\label{sec:charged-demianski}

In this section we solve Einstein–Maxwell equations \eqref{top-down:eq:einstein-maxwell-eom} in the particular case where
\begin{equation}
	\label{eq:static-ansatz-one-unknown}
	f_t = f, \qquad
	f_r = f^{-1}, \qquad
	f_\Omega = r^2.
\end{equation} 

First the static solution is recalled for later comparison.
The stationary solution is derived in $(u, r)$ coordinates in order to avoid the question of the validity of the Boyer–Lindquist transformation.

\subsection{Static case}

Consider the static metric \eqref{top-down:eq:static-metric} and gauge field \eqref{top-down:eq:static-gauge-field}.

Only the $(t)$ component of Maxwell equations is non trivial
\begin{equation}
	2 f'_A + r f''_A = 0,
\end{equation} 
the prime being a derivative with respect to $r$, and its solution is
\begin{equation}
	f_A(r) = \alpha + \frac{q}{r}
\end{equation} 
where $q$ is a constant of integration that is interpreted as the charge and $\alpha$ is an additional constant that can be removed by a gauge transformation.

The only relevant Einstein equation is
\begin{equation}
	\frac{q^2}{r^2} - \kappa + r^2 \Lambda + f + r f' = 0
\end{equation} 
whose solution reads
\begin{equation}
	\label{eq:topdown-1:static-f}
	f(r) = \kappa - \frac{2m}{r} + \frac{q^2}{r^2} - \frac{\Lambda}{3}\, r^2,
\end{equation} 
$m$ being a constant of integration that is identified to the mass.

We stress that we are just looking for solutions of Einstein equations and we are not concerned with regularity (in particular it is well-known that only $\kappa = 1$ is well-defined for $\Lambda = 0$).

\subsection{Stationary case}

We turn our attention to the Einstein–Maxwell equations \eqref{top-down:eq:einstein-maxwell-eom} for the stationary metric \eqref{top-down:eq:rotating-metric:ur} and gauge field \eqref{top-down:eq:rotating-gauge-field:ur} obtained from the seed metric \eqref{eq:static-ansatz-one-unknown} through the DJN algorithm.

\subsubsection{Simplifying the equations}
\label{sec:charged-demianski:simplifying}

The components $(rr)$ and $(r\theta)$ give respectively the equation
\begin{subequations}
\begin{align}
	G'' + \frac{H'}{H}\; G' &= \pm 2 F, \\
	F' \left( G'' + \frac{H'}{H}\; G' \right) &= 2 F F'.
\end{align}
\end{subequations}
\footnotetext{In particular all expressions are quadratic in $F$, but only linear in $F'$.}
On the other hand if $F' \neq 0$ one can simplify by the latter in the second equation and this fixes the sign of the first equation.
Then in both cases the relevant equation reduces to
\begin{equation}
	\label{eq:topdown-1-F-Gd-bis}
	G'' + \frac{H'}{H}\; G' = 2 F,
\end{equation} 
which depends only on $\theta$ and allows to solve for $G$ in terms of $F$.

Integrating the $r$-component of the Maxwell equation gives
\begin{equation}
	\tilde f_A = \frac{q\, r}{r^2 + F^2} + \alpha\, \frac{r^2 - F^2}{r^2 + F^2}.
\end{equation}
The $\theta$-equation reads
\begin{equation}
	\alpha\, F' = 0
\end{equation}
which implies $\alpha = 0$ if $F' \neq 0$.
The $\phi$- and $t$-equations follow from these two equations.
As seen above, $\alpha$ can be removed in the static limit $F \to 0$ and in the rest of this section we consider only the case\footnotemark{}
\begin{equation}
	\alpha = 0.
\end{equation} 
\footnotetext{We relax this assumption in appendix~\ref{app:relaxing-assumptions:gauge-fields}.}

The $(tr)$ equation contains only $r$-derivatives of $\tilde f$ and can be integrated, giving\footnotemark{}
\footnotetext{In~\cite{Demianski:1972:NewKerrlikeSpacetime} the last term of $\tilde f$ is missing as pointed out in~\cite{Quevedo:1992:ComplexTransformationsCurvature}, as can be compared with other references on (A)dS–Taub–NUT, see for example~\cite{AlonsoAlberca:2000:SupersymmetryTopologicalKerrNewmannTaubNUTaDS}.}
\begin{equation}
	\tilde f = \kappa - \frac{2m r - q^2 + 2 F (\kappa\, F + K)}{r^2 + F^2} - \frac{\Lambda}{3}\, (r^2 + F^2) - \frac{4 \Lambda}{3}\, F^2 + \frac{8 \Lambda}{3}\, \frac{F^4}{r^2 + F^2}
\end{equation} 
where again $m$ is a constant of integration interpreted as the mass, and the function $K$ is defined by
\begin{equation}
	2 K = F'' + \frac{H'}{H}\, F'.
\end{equation} 
This implies the equations $(r\phi)$ and $(\theta\theta)$.

As explained in section~\ref{sec:generalized-jna:jna-transf} the $\theta$-dependence should be contain in $F(\theta)$ only.
The second term of the function $\tilde f$ contains some lonely $\theta$ from the $H(\theta)$ in the function $K$: this means that they should be compensated by the $F$, and we therefore ask that the sum $\kappa F + K$ be constant\footnotemark{}
\footnotetext{In appendix~\ref{app:relaxing-assumptions:metric-function} we relax this last assumption by allowing non-constant $\kappa F + K$.
In this context the equations and the function $\tilde f$ are modified and this provides an explanation for the error in $\tilde f$ of Demiański's paper~\cite{Demianski:1972:NewKerrlikeSpacetime}.}
\begin{equation}
	\kappa\, F' + K' = 0
	\Longrightarrow
	\kappa\, F + K = \kappa\, n.
\end{equation} 
The parameter $n$ is interpreted as the NUT charge.

The components $(t\theta)$ and $(\theta\phi)$ give the same equation
\begin{equation}
	\Lambda\, F' = 0.
\end{equation} 

Finally one can check that the last three equations $(tt), (t\phi)$ and $(\phi\phi)$ are satisfied.

Let's summarize the equations
\begin{subequations}
\label{eq:topdown-1}
\begin{align}
	\label{eq:topdown-1-F-Gd}
	2 F &= G'' + \frac{H'}{H}\; G', \\
	\label{eq:topdown-1-Fd-Kd}
	\kappa\, n &= \kappa\, F + K, \\
	\label{eq:topdown-1-lambda}
	0 &= \Lambda F'
\end{align}
and the function $\tilde f$
\begin{equation}
	\label{eq:topdown-1-tilde-f}
	\tilde f = \kappa - \frac{2m r - q^2 + 2 F (\kappa\, F + K)}{r^2 + F^2} - \frac{\Lambda}{3}\, (r^2 + F^2) - \frac{4 \Lambda}{3}\, F^2 + \frac{8 \Lambda}{3}\, \frac{F^4}{r^2 + F^2}.
\end{equation}
We also defined
\begin{equation}
	\label{eq:topdown-1-K-Fd}
	2 K = F'' + \frac{H'}{H}\, F'.
\end{equation} 
\end{subequations}

As explained in the introduction, a major issue of Demiański's approach is the impossibility to obtain – at least in a direct manner – the stationary $\tilde f$ function \eqref{eq:topdown-1-tilde-f} as a complexification of the static $f$ function \eqref{eq:topdown-1:static-f}.
This is one of the reason explaining why applications of the JN algorithm have been limited to adding a rotation parameter until now.
We address this question in section~\ref{sec:complexification} and show how to recover $\tilde f$ from $f$.

In the next sections we solve explicitly the equations \eqref{eq:topdown-1} in both cases $\Lambda \neq 0$ and $\Lambda = 0$.

\subsubsection{Solution for \texorpdfstring{$\Lambda \neq 0$}{non-vanishing cosmological constant}}

Equation \eqref{eq:topdown-1-lambda} implies that $F' = 0$, from which $K = 0$ follows by definition; then one obtains
\begin{equation}
	F(\theta) = n
\end{equation} 
by compatibility with \eqref{eq:topdown-1-Fd-Kd} and since $K(\theta) = 0$.

Solution to \eqref{eq:topdown-1-F-Gd} is
\begin{equation}
	G(\theta) = c_1 - 2 \kappa\, n \ln H(\theta) + c_2 \ln \frac{H(\theta/2)}{H'(\theta/2)}
\end{equation} 
where $c_1$ and $c_2$ are two constants of integration.
Since only $G'$ appears in the metric we can set $c_1 = 0$.
On the other hand the constant $c_2$ can be removed by the transformation
\begin{equation}
	\dd u = \dd u' - c_2\, \dd\phi.
\end{equation} 

We summarize the solution to the system \eqref{eq:topdown-1}
\begin{equation}
	F(\theta) = n, \qquad
	G(\theta) = - 2 \kappa\, n \ln H(\theta).
\end{equation} 

The function $\tilde f$ then takes the form
\begin{equation}
	\label{eq:topdown-1:tilde-f-lambda}
	\tilde f = \kappa - \frac{2m r - q^2 + 2 \kappa n^2}{r^2 + n^2} - \frac{\Lambda}{3}\,\frac{r^4 + 6 n^2 r^2 - 3 n^4}{r^2 + n^2}.
\end{equation} 

The transformation to BL coordinates is well defined
\begin{equation}
	g = \frac{r^2 + n^2}{\Delta}, \qquad
	h = 0, \qquad
	\Delta = \kappa r^2 - 2 m r + q^2 + \Lambda n^4 - \frac{\Lambda}{3}\, r^4 - n^2 (\kappa + 2 \Lambda r^2 )
\end{equation} 
in the sense that none of these functions depend on $\theta$, as recalled in section~\ref{sec:generalized-jna:jna-transf}.

As noted by Demiański the only parameters that appear are the mass and the NUT charge, and it is not possible to add an angular momentum for non-vanishing cosmological constant.\footnotemark{}
As a consequence the DJN algorithm cannot provide a derivation of (A)dS–Kerr–Newman.
\footnotetext{In~\cite{Leigh:2014:GerochGroupEinstein} Leigh et al.\ generalized Geroch's solution generating technique and also found that only the mass and the NUT charge appear when $\Lambda \neq 0$. We would like to thank D.\ Klemm for this remark.}

\subsubsection{Solution for \texorpdfstring{$\Lambda = 0$}{vanishing cosmological constant}}

The solution to the differential equation \eqref{eq:topdown-1-Fd-Kd} is
\begin{equation}
	F(\theta) = n - a\, H'(\theta) + c \left( 1 + H'(\theta)\, \ln \frac{H(\theta/2)}{H'(\theta/2)} \right)
\end{equation}
where $a$ and $c$ denote two constants of integration.

We solve the equation \eqref{eq:topdown-1-F-Gd} for $G$
\begin{equation}
	\begin{aligned}
		G(\theta) = c_1 &+ \kappa\, a\, H'(\theta)
			- \kappa\, c\, H'(\theta)\, \ln \frac{H(\theta/2)}{H'(\theta/2)}
			- 2 \kappa\, n \ln H(\theta) \\
			&+ (a + c_2) \ln \frac{H(\theta/2)}{H'(\theta/2)}
	\end{aligned}
\end{equation} 
and $c_1, c_2$ are constants of integration.
Again since only $G'$ appears in the metric we can set $c_1 = 0$.
We can also remove the last term with the transformation
\begin{equation}
	\dd u = \dd u' - (c_2 + a) \dd\phi.
\end{equation} 

One finally gets
\begin{subequations}
\begin{align}
	F(\theta) &= n - a\, H'(\theta) + c \left( 1 + H'(\theta)\, \ln \frac{H(\theta/2)}{H'(\theta/2)} \right), \\
	G(\theta) &= \kappa\, a\, H'(\theta)
		- \kappa\, c\, H'(\theta)\, \ln \frac{H(\theta/2)}{H'(\theta/2)}
		- 2 \kappa\, n \ln H(\theta).
\end{align}
\end{subequations}

The Boyer–Lindquist transformation is well defined only for $c = 0$, in which case
\begin{equation}
	g = \frac{r^2 + a^2 + n^2}{\Delta}, \qquad
	h = \frac{\kappa a}{\Delta}, \qquad
	\Delta = \kappa r^2 - 2 m r + q^2 - \kappa n^2 + \kappa a^2.
\end{equation} 
The function $\tilde f$ reads~\cite[sec.~2.2]{AlonsoAlberca:2000:SupersymmetryTopologicalKerrNewmannTaubNUTaDS}
\begin{equation}
	\label{eq:topdown-1:tilde-f-no-Lambda-no-c}
	\tilde f = \kappa - \frac{2 m r - q^2}{\rho^2} + \frac{\kappa\, n (n - a H')}{\rho^2}, \qquad
	\rho^2 = r^2 + (n - a\, H')^2.
\end{equation} 

The constant $a$ corresponds to the angular momentum (and one recognizes the usual JN algorithm), while $c$ is not easy to interpret~\cites{Adamo:2014:KerrNewmanMetricReview}[sec.~5.3]{Krasinski:2006:InhomogeneousCosmologicalModels}.

This solution was already found in~\cite{Krori:1981:ChargedDemianskiMetric} for the case $\kappa = 1$ by solving directly Einstein–Maxwell equations, starting with a metric ansatz of Demiański's form.
In our case we aim to show that the same solution can be obtained by applying Demiański's method to all the quantities, including the gauge field.

\section{Group properties}
\label{sec:group-properties}

In this section we want to study the DJN transformations that form a group.
The main motivation is to state clearly when several transformations can be chained, for example when one starts with a solution that already has a parameter that could be added with the DJN algorithm.

We will make the assumptions that the functions $F(\theta)$ and $G(\theta)$ are linear in some parameters $\pi^A$ (implicit sum over $A$)
\begin{equation}
	F(\theta) = \pi^A F_A(\theta), \qquad
	G(\theta) = \pi^A G_A(\theta),
\end{equation} 
where $\{ F_A(\theta) \}$ and $\{ G_A(\theta) \}$ are the functions associated to the parameters and $A$ runs over the dimension of this space.
Mathematically the functions are member of an additive group $\mc G$ with elements in\footnotemark{} $\mc F \times \mc F$ ($\mc F$ being the space of functions with second derivatives) with generators $\big( F_A(\theta), G_A(\theta) \big)$, $A = 1, \ldots, \dim \mc V$ since there is an identity element $0$ and each element with coefficients $\pi^A$ possesses an inverse given by $- \pi^A$.
Adding the multiplication by a scalar turns this group into a vector space but we do not need this extra structure.
As a consequence the sum of two functions $F_1 = \pi_1^A F_A$ and $F_2 = \pi_2^A F_A$ gives another function $F_3 = \pi_3^A F_A$ with $\pi_3^A = \pi_1^A + \pi_2^A$.
These assumptions are motivated by the results of section~\ref{sec:charged-demianski} where $F$ and $G$ were solutions of (non-homogeneous) second order linear differential equations where the $\pi^A$ are the integration constants.
\footnotetext{For simplicity we consider the case where $F$ and $G$ are expanded over the same parameters, but this is not necessarily the case.}

After a first transformation
\begin{equation}
	r = r' + i\, F_1, \qquad
	u = u' + i\, G_1
\end{equation} 
one obtains the metric (omitting the primes)
\begin{equation}
	\begin{aligned}
		\dd s^2 = &- \tilde f^{\{1\}}_t (\dd u + H G_1'\, \dd\phi)^2
			+ \tilde f^{\{1\}}_\Omega (\dd\theta^2 + H^2 \dd\phi^2) \\
			&- 2 \sqrt{\tilde f^{\{1\}}_t \tilde f^{\{1\}}_r} (\dd u + G_1' H\, \dd\phi) (\dd r + F_1' H\, \dd\phi)
	\end{aligned}
\end{equation} 
where
\begin{equation}
	\tilde f^{\{1\}}_i = \tilde f^{\{1\}}_i(r, F_1).
\end{equation} 
Performing a second transformation
\begin{equation}
	r = r' + i\, F_2, \qquad
	u = u' + i\, G_2
\end{equation} 
the previous metric becomes (omitting the primes)
\begin{equation}
	\label{topdown:eq:metric-two-transf}
	\begin{aligned}
		\dd s^2 = &- \tilde f^{\{1,2\}}_t \big( \dd u + H (G_1' + G_2')\, \dd\phi \big)^2
			+ \tilde f^{\{1,2\}}_\Omega (\dd\theta^2 + H^2 \dd\phi^2) \\
			&- 2 \sqrt{\tilde f^{\{1,2\}}_t \tilde f^{\{1,2\}}_r} \big( \dd u + (G_1' + G_2') H\, \dd\phi \big) \big( \dd r + (F_1' + F_2') H\, \dd\phi \big)
	\end{aligned}
\end{equation} 
where
\begin{equation}
	\tilde f^{\{1,2\}}_i = \tilde f^{\{1,2\}}_i(r, F_1, F_2).
\end{equation}
This function is required to satisfy the following conditions (omitting the primes)
\begin{equation}
	\tilde f^{\{1,2\}}_i(r, F_1, 0) = \tilde f^{\{1\}}_i(r, F_1), \qquad
	\tilde f^{\{1,2\}}_i(r, F_1, F_2) = \tilde f^{\{2,1\}}_i(r, F_2, F_1).
\end{equation} 
The second condition means that the order of the transformations should not matter because we want to obtain the same solution given identical seed metric and parameters.

The metric \eqref{topdown:eq:metric-two-transf} is obviously equivalent to the one we would get with a unique transformation\footnotemark{}
\begin{equation}
	r = r' + i\, (F_1 + F_2), \qquad
	u = u' + i\, (G_1 + G_2).
\end{equation} 
\footnotetext{This breaks down when the metric is transformed with more complicated rules, such as in higher dimensions~\cite{Erbin:2015:FivedimensionalJanisNewmanAlgorithm}.}
Then, for the transformations which are such that
\begin{equation}
	\label{topdown:eq:fi-sum-F}
	\tilde f^{\{1,2\}}_i(r, F_1, F_2) = \tilde f^{\{1\}}_i(r, F_1 + F_2),
\end{equation} 
the DJN transformations form an Abelian group thanks to the group properties of the function space.
This structure implies that we can first add one parameter, and later another one (say first the NUT charge, and then an angular momentum).
Said another way this group \emph{preserves Einstein equations} when the seed metric is a known (stationary) solution.
But note that it may be very difficult to do it as soon as one begins to replace the $F$ in the functions by their expression, because it obscures the original function – in one word we can not find $\tilde f_i(r, F)$ from $\tilde f_i(r, \theta)$.

Another point worth to mention is that not all DJN transformation are in this group since the condition \eqref{topdown:eq:fi-sum-F} may not satisfied: we recall that imposing or not the latter is a choice that one is doing when performing the algorithm.
A simple example is provided by
\begin{equation}
	f(r) = r^2,
\end{equation} 
which can be transformed under the two successive transformations
\begin{equation}
	r = r' + i F_1, \qquad
	r' = r'' + i F_2
\end{equation} 
in two ways:
\begin{subequations}
\begin{align}
	1.& \qquad
		\tilde f^{\{1\}} = \abs{r}^2
			= r'^2 + F_1^2, \qquad
		\tilde f^{\{1,2\}} = \abs{r'}^2 + F_1^2
			= r''^2 + F_1^2 + F_2^2, \\
	2.& \qquad
		\tilde f^{\{1\}} = \abs{r}^2
			= \abs{r' + i F_1}^2, \qquad
		\begin{aligned}
			\tilde f^{\{1,2\}} &= \abs{r'' + i (F_1 + F_2)}^2 \\
				&= r''^2 + F_1^2 + F_2^2 + 2 F_1 F_2.
		\end{aligned}
\end{align}
\end{subequations}
Only the second option satisfy the property \eqref{topdown:eq:fi-sum-F} that leads to a group.
Such an example is provided in $5d$ where the function $f_\Omega(r) = r^2$ is successively transformed as~\cite{Erbin:2015:FivedimensionalJanisNewmanAlgorithm}
\begin{equation}
	r^2 \longrightarrow \abs{r}^2 = r^2 + a^2 \cos^2 \theta \longrightarrow \abs{r}^2 + a^2 \cos^2 \theta = r^2 + a^2 \cos^2 \theta + b^2 \sin^2 \theta,
\end{equation} 
with the functions
\begin{equation}
	F_1 = a \cos \theta, \qquad
	F_2 = b \sin \theta.
\end{equation} 
The condition \eqref{topdown:eq:fi-sum-F} is clearly not satisfied.

\section{Finding the complexification}
\label{sec:complexification}

At the end of section~\ref{sec:charged-demianski:simplifying}, we mentioned the issue of finding the complexification of the stationary function from the static one.
This is a key step if one wishes to apply the algorithm – with the new parameters $n$ and $c$ – to other cases.

When one considers only rotation a set of rules has been established~\cite{Newman:1965:NoteKerrSpinningParticle, Newman:1965:MetricRotatingCharged}
\begin{subequations}
\label{eq:complexification-rules}
\begin{align}
	\label{eq:complexification-rules-r}
	r & \longrightarrow \frac{1}{2} (r + \bar r) = \Re r\,, \\
	\label{eq:complexification-rules-1/r}
	\frac{1}{r} & \longrightarrow \frac{1}{2} \left(\frac{1}{r} + \frac{1}{\bar r}\right) = \frac{\Re r}{\abs{r}^2}\,, \\
	\label{eq:complexification-rules-r2}
	r^2 & \longrightarrow \abs{r}^2.
\end{align}
\end{subequations}
Such transformations are still valid while adding the parameter $c$ but are not sufficient when one is considering the NUT charge $n$.
Indeed the last case also requires the complexification of the mass parameter.
In what follows we ignore the electric charge since it does not modify the discussion.\footnotemark{}
\footnotetext{Indeed the electric charge appears only as a term $q^2 / r^2$ which is decoupled from the NUT charge.}

\subsection{\texorpdfstring{$\Lambda = 0$}{Vanishing cosmological constant}}

The static Schwarzschild function \eqref{eq:topdown-1:static-f}
\begin{equation}
	f = \kappa - \frac{2m}{r}
\end{equation} 
is complexified as
\begin{equation}
	\tilde f = \kappa - \left(\frac{m}{r} + \frac{\bar m}{\bar r} \right)
		= \kappa - \frac{2 \Re(m \bar r)}{\abs{r}^2}.
\end{equation} 

Performing the transformation
\begin{equation}
	\label{eq:complex-mass-no-lambda}
	m = m' + i \kappa\, n, \qquad
	r = r' + i F
\end{equation} 
gives
\begin{equation}
	\tilde f = \kappa - \frac{2 m r + 2 \kappa n F}{r^2 + F^2}
\end{equation}
which corresponds to the correct function \eqref{eq:topdown-1-tilde-f}.

\subsection{\texorpdfstring{$\Lambda \neq 0$}{Non-vanishing cosmological constant}}

The static Schwarzschild function \eqref{eq:topdown-1:static-f}
\begin{equation}
	f = \kappa - \frac{2m}{r} - \frac{\Lambda}{3}\, r^2
\end{equation} 
is complexified as
\begin{equation}
	\tilde f = \kappa - \frac{2 \Re(m \bar r)}{\abs{r}^2} - \frac{\Lambda}{3}\, \abs{r}^2.
\end{equation} 

The transformation of the function $\tilde f$ involves a complex transformation of the mass together with a shift of the horizon curvature\footnotemark{}
\footnotetext{Notice that AdS–Taub–NUT (for $\kappa = -1$, $m = 0$) is supersymmetric for $n = \pm 1/(2g)$ where $g^2 = - \Lambda / 3$~\cite[tab.~1]{AlonsoAlberca:2000:SupersymmetryTopologicalKerrNewmannTaubNUTaDS}.}
\begin{equation}
	\label{eq:complex-mass-lambda}
	m = m' + i \kappa\, n, \qquad
	r = r' + i n, \qquad
	\kappa = \kappa' - \frac{4\Lambda}{3}\, n^2.
\end{equation}
One needs to accept the fact that $\kappa$ and $\kappa'$ take the same values $\pm 1$ even if this is not true in the previous formula; this may be interpreted as a rescaling of the horizon curvature $\kappa$ due to the coupling of the cosmological constant with the NUT charge (recall that the values $\kappa = \pm 1$ correspond to a choice of scale).
One can notice that the limit $\Lambda \to 0$ agrees with the previous section (upon replacing $n$ by $F$).
In total the mass is transformed as\footnotemark{}
\begin{equation}
	m = m' + i n \left( \kappa' - \frac{4\Lambda}{3}\, n^2 \right).
\end{equation} 
\footnotetext{The imaginary part of the new mass term appears in other contexts~\cite{Plebanski:1975:ClassSolutionsEinsteinMaxwell, Quevedo:1992:DeterminationMetricCurvature, Chamblin:1999:LargeNPhases, Johnson:2014:ThermodynamicVolumesAdSTaubNUT}.
In particular this corresponds to a condition of regularity in Euclidean signature.}

Inserting these transformations into $\tilde f$ gives the result
\begin{equation}
	\tilde f = \kappa - \frac{2 m r + 2 \kappa n^2}{r^2 + n^2} - \frac{\Lambda}{3}\, (r^2 + 5 n^2) + \frac{8 \Lambda}{3}\,\frac{n^4}{r^2 + n^2}
\end{equation} 
and we retrieve \eqref{eq:topdown-1:tilde-f-lambda}.

\section{Conclusion and discussion}

In this paper we generalize Demiański analysis of the JN algorithm.
Our main result consists in the proof that Demiański's solution~\cite{Demianski:1972:NewKerrlikeSpacetime} can be recovered by a direct application of the JN prescription, and not only by solving Einstein equations.
As a consequence it becomes possible to perform Demiański's transformations on other seed metrics with and without cosmological constant.

Furthermore we showed that this solution can be generalized to include an electric charge and also to possess a topological horizon.
Hence the final metric contains (for vanishing cosmological constant) four of the six Plebański–Demiański parameters~\cite{Plebanski:1976:RotatingChargedUniformly} along with Demiański's parameter.
Having complex parameters is a key step for adding a magnetic charge and for getting supergravity solutions, and we reserve this for later work.
Note that, intriguingly, one could het here all Plebański–Demiański parameters but the acceleration.

Demiański derived his transformation by solving Einstein equations in the presence of only one unknown function.
We were not able to solve analytically Einstein–Maxwell equations for a greater number of unknown functions, but looking at the equations indicate that the same structure persists in these more general cases.
Therefore we can hope that the transformations obtained in section~\ref{sec:charged-demianski} are the most general ones under the assumptions of our paper.
This is also motivated by the examples from~\cite{Erbin:2015:SupergravityComplexParameters} and by the solutions in~\cite{Krori:1981:ChargedDemianskiMetric, Patel:1988:RadiatingDemianskitypeMetrics}.

A result from Demiański's analysis is the impossibility to find Kerr–AdS from the DJN algorithm and it is often quoted as a no-go theorem.
But this outcome relies on the assumption that no parameter already present in the static metric is complexified, which may be a too strong statement.

Another advantage of Demiański's approach resides in the fact that it can be used to find the form of the transformation leading to solutions without ambiguity.
In particular this approach could be used to generalize the results of~\cite{Erbin:2015:FivedimensionalJanisNewmanAlgorithm}.

Finally Drake and Szekeres obtained general results on the solutions that can be derived from the original JN algorithm~\cite{Drake:2000:UniquenessNewmanJanisAlgorithm}.
In their work they invert the Boyer–Lindquist equations in order to get the metric functions in terms of the BL functions, which are then obtained from Einstein equations.\footnotemark{}
This is another way to bypass the complexification and obtain general information on its properties, and it would be very interesting to extend their analysis with the more general Demiański's transformation.
\footnotetext{This idea has been generalized in~\cite{AzregAinou:2014:GeneratingRotatingRegular} to include a third function.}

\section*{Acknowledgments}

I wish to thank Lucien Heurtier for many discussions and collaborations on related topics and for corrections on the draft version.
I am also very grateful to Nick Halmagyi and Dietmar Klemm for support and interesting discussions, and to Corinne de Lacroix for reading the manuscript.

This work, made within the \textsc{Labex Ilp} (reference \textsc{Anr–10–Labx–63}), was supported by French state funds managed by the \emph{Agence nationale de la recherche}, as part of the programme \emph{Investissements d'avenir} under the reference \textsc{Anr–11–Idex–0004–02}.

\appendix

\section{Original Demiański's solution}
\label{app:original-demianski}

In this appendix we recall the Demiański's result, amended to fix some errors~\cite{Quevedo:1992:ComplexTransformationsCurvature}.
They follow from section~\ref{sec:charged-demianski} with $\kappa = 1$ and $q = 0$.
This gives also the opportunity to present prettier formulas.

The equations are
\begin{subequations}
\begin{align}
	2 F &= G'' + \cot \theta\; G', \\
	n &= F + K, \\
	0 &= \Lambda F'
\end{align}
\end{subequations}
with
\begin{equation}
	2 K = F'' + \cot \theta\, F'.
\end{equation} 
The function $\tilde f$ is
\begin{equation}
	\tilde f = 1 - \frac{2m r - q^2 + 2 F (F + K)}{r^2 + F^2} - \frac{\Lambda}{3}\, (r^2 + F^2) - \frac{4 \Lambda}{3}\, F^2 + \frac{8 \Lambda}{3}\, \frac{F^4}{r^2 + F^2}.
\end{equation}

The two solutions for $F$ and $G$ are
\begin{itemize}
	\item $\Lambda \neq 0$
	\begin{equation}
		F(\theta) = n, \qquad
		G(\theta) = - 2 n \ln \sin(\theta).
	\end{equation} 
	
	\item $\Lambda = 0$
	\begin{subequations}
	\begin{align}
		F &= n - a \cos \theta + c \left( 1 + \cos \theta\, \ln \tan \frac{\theta}{2} \right), \\
		G &= a \cos \theta - 2 n \ln \sin \theta - c \cos \theta\, \ln \tan \frac{\theta}{2}.
	\end{align}
	\end{subequations}
\end{itemize}

\section{Relaxing assumptions}
\label{app:relaxing-assumptions}

\subsection{Metric function \texorpdfstring{$F$}{F}-dependence}
\label{app:relaxing-assumptions:metric-function}

In section~\ref{sec:charged-demianski:simplifying} we obtained the equation \eqref{eq:topdown-1-Fd-Kd}
\begin{equation}
	\kappa\, F + K = \kappa\, n, \qquad
	2 K = F'' + \frac{H'}{H}\, F'
\end{equation}
by asking that the function \eqref{eq:topdown-1-tilde-f}
\begin{equation}
	\tilde f = \kappa - \frac{2m r - q^2 + 2 F (\kappa\, F + K)}{r^2 + F^2} - \frac{\Lambda}{3}\, (r^2 + F^2) - \frac{4 \Lambda}{3}\, F^2 + \frac{8 \Lambda}{3}\, \frac{F^4}{r^2 + F^2}
\end{equation} 
depends on $\theta$ only through $F(\theta)$.

A more general assumption would be that $\kappa F + K$ is some function $\chi = \chi(F)$
\begin{equation}
	\label{eq:topdown-1-Fd-Kd-chi}
	\kappa\, F + K = \kappa\, \chi(F).
\end{equation} 
First if $F' = 0$ then $K = 0$ and the definition of $K$ implies
\begin{equation}
	\chi = F = n.
\end{equation} 
The $(t\theta)$- and $(\theta\phi)$-components give the equation
\begin{equation}
	4 \Lambda\, F^2 F' = F'\, \pd_F \chi.
\end{equation} 

If $\Lambda = 0$ we find that
\begin{equation}
	\pd_F \chi = 0
	\Longrightarrow
	\chi = n
\end{equation} 
which reduces to the case studied in section~\ref{sec:charged-demianski:simplifying}, while if $F' = 0$ this equation does not provide anything.

On the other hand if $F' \neq 0$ and $\Lambda \neq 0$ then the previous equation becomes
\begin{equation}
	\pd_F \chi = 4 \Lambda F^2
\end{equation} 
which can be integrated to
\begin{equation}
	\label{top-down:eq:chi-F-solution}
	\chi(F) = n + \frac{4}{3}\, \Lambda F^3
\end{equation} 
(notice that the limit $\Lambda \to 0$ is coherent).
Plugging this function into equation \eqref{eq:topdown-1-Fd-Kd-chi} one obtains
\begin{equation}
	\label{eq:topdown-1-Fd-Kd-chi-replaced}
	\kappa\, F + K = \kappa \left(n + \frac{4}{3}\, \Lambda F^3 \right)
\end{equation} 
(remember that $F' \neq 0$).
This differential equation is non-linear and we were not able to find an analytical solution.
Despite that this provides a generalization of the algorithm with non-constant $F$ in the presence of a cosmological constant this is not sufficient for obtaining Kerr–(a)dS: the form of $g_{\theta\theta}$ given in \eqref{top-down:eq:rotating-metric:tr} is not required one.

Nonetheless by inserting the expression of $\chi$ in $\tilde f$ we see that the last term is killed
\begin{equation}
	\tilde f = \kappa - \frac{2m r - q^2 + 2 \kappa\, n\, F}{r^2 + F^2} - \frac{\Lambda}{3}\, (r^2 + F^2) - \frac{4 \Lambda}{3}\, F^2.
\end{equation} 
One can recognize the function given by Demiański~\cite{Demianski:1972:NewKerrlikeSpacetime}.
Then this function is valid at the condition that equation \eqref{eq:topdown-1-Fd-Kd} is modified to \eqref{eq:topdown-1-Fd-Kd-chi-replaced}, but in this case the solution is not the general (A)dS–Taub–NUT anymore.

\subsection{Gauge field integration constant}
\label{app:relaxing-assumptions:gauge-fields}

In section~\ref{sec:charged-demianski:simplifying} we obtained a second integration constant $\alpha$ in the expression of the gauge field
\begin{equation}
	\tilde f_A = \frac{q\, r}{r^2 + F^2} + \alpha\, \frac{r^2 - F^2}{r^2 + F^2}.
\end{equation}
One of the Maxwell equation gave $\alpha = 0$ if $F' \neq 0$, but otherwise no equation fixes its value.
For this reason we focus on the case $F' = 0$ or equivalently $\Lambda = 0$ through equation \eqref{eq:topdown-1-lambda}.

In this case the function $\tilde f$ is modified to
\begin{equation}
	\tilde f = \kappa - \frac{2m r - q^2 + 2 F (\kappa\, F + K) + 4 \alpha^2 F^2}{r^2 + F^2} - \frac{\Lambda}{3}\, (r^2 + F^2) - \frac{4 \Lambda}{3}\, F^2 + \frac{8 \Lambda}{3}\, \frac{F^4}{r^2 + F^2}.
\end{equation} 
Equation \eqref{eq:topdown-1-lambda} is modified but it is still solved by $F' = 0$ and all other equations are left unchanged (in particular $\kappa F + K$ is still given by the function $\chi(F)$ \eqref{top-down:eq:chi-F-solution}).
For $\chi(F) = n$ the configuration with $\alpha \neq 0$ provides another solution when $\Lambda \neq 0$ but it is not clear how to get it from a complexification of the function.

\printbibliography[heading=bibintoc]

\end{document}